# DETECTION, GROWTH QUANTIFICATION AND MALIGNANCY PREDICTION OF PULMONARY NODULES USING DEEP CONVOLUTIONAL NETWORKS IN FOLLOW-UP CT SCANS

Xavier Rafael-Palou[a,b], Anton Aubanell[c], Mario Ceresa[b], Vicent Ribas[a], Gemma Piella[b] and Miguel A. González Ballester[b,d]

[a]Eurecat Centre Tecnològic de Catalunya, eHealth Unit, Barcelona, Spain
[b]BCN MedTech, Dept. of Information and Communication Technologies, Universitat Pompeu Fabra, Barcelona, Spain
[c]Vall d'Hebron University Hospital, Barcelona, Spain
[d]ICREA, Barcelona, Spain

## 1. Introduction

The use of computed tomography (CT) scan images has increased dramatically over the last decades, becoming a crucial tool for the diagnosis and follow-up of malignant lung tumours [1,2]. Radiologists are able to detect, measure and monitor the evolution of abnormal tissues in their lungs by visually inspecting CT scans of the patient's chest. However, tumours, specially at early stages, are complex to detect and diagnose due to large heterogeneities in their morphology, size, texture, localization and growth rates [3]. Moreover, spatial resolution in computerized axial tomography images is often limited by the acquisition protocol [4]. This leads to some ambiguities and conflicts for radiologists when having to determine the next study, whether to discharge the patient from the follow-up, or whether to resolve a clinical intervention for the patient [5]. Therefore, the experience and expertise of physicians is fundamental for the early diagnosis and prognosis of lung cancer. Unfortunately, the aggressive nature of this disease, its important incidence in the adult population, and the constant need of specialized professionals, make it necessary to have accurate and efficient tools to reduce the workload of clinicians as well as to help them in making critical decisions.

The idea of providing automatic support for the detection and diagnosis of lung cancer is not new, and large efforts have been made with conventional machine learning and artificial intelligence techniques [6-8]. Recently, the advent of deep neural networks [9] has allowed a major breakthrough in the medical image domain [10-12]. Specifically for lung cancer, outstanding performances have been achieved in a very short period of time, outperforming conventional approaches for nodule detection [13], pixel segmentation [14], or lung cancer classification [15]. Despite this, most of the research focuses primarily on a single CT scan. This fact conditions the potential of these contributions, since they do not consider the temporal evolution of the tumour, which, indeed, is one of the most important clinical factors influencing prognosis [31].

In this work we take a step forward in supporting the radiological workflow, by proposing an automatic tool that takes into account the evolution of the pulmonary nodules in the predictive modeling task. To do this, we defined a data-driven approach with a flexible and configurable four-stage pipeline, which 1) automatically detects nodules, 2) re-identifies them from different CT scans of a given patient, 3) quantifies their growth, and 4) predicts their malignancy. To configure each of the pipeline components, we have integrated existing solutions [16-18] and proposed new ones based on deep convolutional neural networks. Hence, in the remainder of this chapter we describe the background, present the pipeline and its different components, and show the results of its evaluation in a longitudinal cohort of more than 30 patients. We conclude this work by discussing the present solution and establishing future works for the automatic temporal lung nodule assessment.

## 2. Background

In this section we review some of the most relevant and recent works proposed for supporting radiologists in the lung cancer assessment. From the different tasks emcompassed by radiologists in the management of this disease, we focus in the most essential ones, such as nodule detectection, nodule quantification, and lung cancer prediction.

### 2.1 Nodule detection

This task consist on screening the entire lung CT volume, searching for small suspicious regions or nodules (usually between 3 mm to 30 mm) [37]. Nowadays, this problem, as in most of the computer vision research areas, is addressed by convolutional neural networks (CNN) [8] able to extract, without human intervention, accurate feature image representations thanks to their shared-weights architecture and translation invariance characteristics. A common approach for automatic nodule detection consists on dividing the problem in two steps [19,20]: candidate detection and false positive reduction. In the first stage, 2D region proposal networks, such as faster region-based networks (Faster-RCNN) [21], are used to extract suspicious regions of interest from the whole CT scan. In the second stage, these regions are classified as normal tissues or nodules using 3D CNN networks, in which the input are 3D image patches around the center of the nodules. Other recent approaches directly address this problem in a single step [22-24]. They re-adapt region proposal networks with 3D deeper architectures (such as ResNet [35] or DenseNet [36]) to directly predict 3D bounding boxes surrounding the nodules.

## 2.2 Nodule quantification

Another important task for lung cancer assessment is determining the size of the nodule. Currently, radiologists calculate the size of the nodule by visual inspection on the CT scan, locating and measuring the largest diameter (in mm) [38]. Usually this measure is extrapolated to 3D dimensions, by means of mathematical operations [25], to approximate the volume of the tumour. Although this process is simple and fast, it entails significant intra and inter-observer variability in the size of the nodule, which can go up to 3 mm in diameter [26]. Since this variability may negatively impact the disease management, several deep learning solutions have addressed nodule size measurement to support clinicians. Some works [22,23] propose learning the diameter of the tumour by extending the nodule detection network (either in 2D or 3D) with a new output in the network. Other solutions build semantic segmentation networks to automatically determine the pixels of the nodules from which the diameter or volume can later be extracted. One of the most common an successful architectures for segmentation is the U-Net [27]. This type of networks uses a convolutional encoder and decoder backbone, tied at different levels by short-cuts, which allow by-passing high level features of the encoder to the decoder, in order to enhance the image reconstruction task. Several extensions of this architecture can be found, such as its 3D formulation [28] or the incorporation of ResNet-like blocks and a Dice-based loss layer, more suitable for segmentation tasks [29]. A more recent approach, nnU-Net [30] has been successfully applied to a multitude of medical segmentation problems (including pulmonary nodule segmentation). One of the benefits of this approach is the automatic fine-tuning of several configuration parameters to the particular type of images to be segmented.

Despite the high performances reported by U-Net-like networks, they address the segmentation problem from a deterministic point of view. However, due to the inherent ambiguity of the problem (often contours of the nodules are not clearly delimited), it is desirable reporting network uncertainty estimates when predicting the size of the nodules. One way to learn model uncertainty is moving from one-input one-output to one-input multiple-output networks. This change of paradigm has already been tackled in deep neural networks through different approaches. One of the simplest approximations consists in ensembling multiple networks in order to provide multiple opinions [32]. Another approach consists in enabling dropout [33] at inference time in order to provide independent pixel-wise probabilities [34]. Another approximation is by deep generative networks, such as generative adversarial networks [59]. This type of networks try to learn, in an unsupervised manner, a direct mapping from a random noise to an output image. To do this, a generator network creates new valid images (from the random noise) with the intention to fool a discriminator network that evaluates whether an image is valid or fake. An extension of this type of networks are conditional GANs [39], in which the goal is to learn structured outputs conditioned on an input image. To do this, the discriminator receives as input the target image to which conditioning the generator. Similar to cGANs, we can find the conditional variational autoencoders (CVAE) [34]. This type of networks propose learning a multi-dimensional latent space that encodes all possible output images. During training, the latent space distribution defined by the encoder is approximated to a normal distribution to ensure continuity and avoid 'mode collapse' commonly seen in GAN approaches [60]. Also, the random vector sampled from the latent space, together with the target image are passed to the decoder (only during training) in order to generate a new plausible image. A recent work, hierarchical probabilistic U-Net (HPU) [40], has been proposed to cover the gap between the generative ability of producing new structured images of the CVAE, with the accuracy of segmenting images of the U-Net. To do this, during training, a posterior U-Net like network, conditioned on the radiologist ground truth nodule, is added to transfer the latent features to a prior U-Net like network by injection, at different levels of the decoder part of this network.

**2.3 Lung cancer prediction**

To support radiologists in the lung cancer prediction, several works have been proposed relying on 2D and 3D inputs, using different deep learning architectures (e.g. CNN, RNN) [41-43], but mostly relying on single CT scan images (commonly derived from the LIDC dataset [49]). Therefore, very few deep learning works have addressed the temporal evolution of pulmonary nodules to support the clinical decision-making. In [44], an end-to-end deep learning based pipeline was presented for lung cancer prediction using two CT studies per patient (current and previous year). This approach proposes three 3D CNN networks, one for analyzing the lung CT image, another for analyzing nodule patches, and a final one, to provide cancer risk prediction using outcomes from previous two components. In [52], a deep learning approach is proposed for predicting lung cancer risk at 3 years and lung cancer-specific mortality. This study, although not being focused on automatic image analysis, uses a multilayer perceptron to ensemble nodule and non-nodule features associated to lung abnormalities. Like in [44], they use the data from the National Lung Screening Trial (NLST), but results were validated in another longitudinal study [53]. In [17], the authors built an automatic pipeline for lung nodule growth quantification using data from a longitudinal cohort of incidental nodules. In this approach, instead of applying lung CT image registration [58], the authors propose a 3D Siamese CNN to re-identify nodules from different CT studies of the patient. That pipeline also allows nodule growth assessment using the diameters predicted by the nodule detection network.

## 3. Temporal Lung Nodule Assessment

Our pipeline takes as input two images from the same patient at different time-points, identifies the lung nodules, and estimates their malignancy and growth. The pipeline consists of 4 main components (see Figure 1): 1) nodule detection, which is done independently on each image; 2) nodule re-identification, which finds the correspondence between nodules across time points; 3) nodule malignancy classification; and 4) nodule growth quantification.

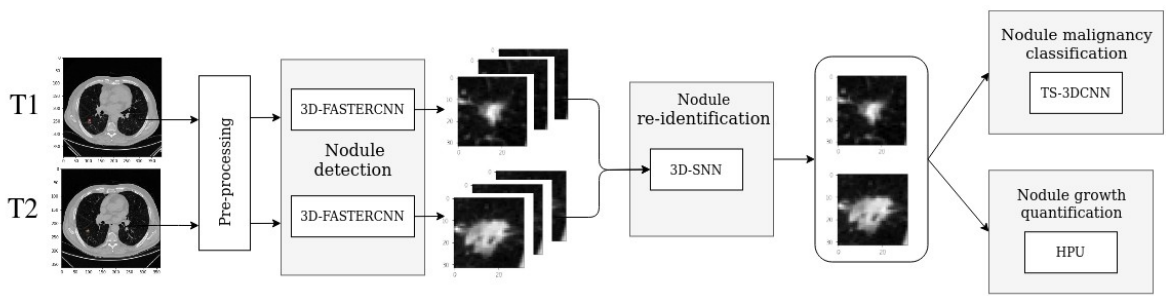

Figure 1. Pipeline architecture for the temporal analysis of lung nodules

The following subsections describe each component, making special emphasis on their input and outputs, the core methods and their configuration.

**3.1 Pre-processing**

Lung CT images are usually originated by different scanners and at different image resolutions. Therefore, the pipeline makes an initial pre-processing step with the intention to standardize the input images. Precisely, first, the images are resampled to an isotropic resolution of 1x1x1 mm$^3$. Second, the image pixel intensities are clipped between [-1000, 600] Hounsfield Units to filter out non-tissue related regions. Finally, the pixels are normalized between 0 and 1.

**3.2 Nodule detection**

The first component of the pipeline consists in detecting pulmonary nodules in CT scan images. To this end, we adopted the solution presented in [17], which relied on two previous and successful works [22, 23]. The neural network follows a Faster-RCNN scheme [21] adapted for 3D images of 128x128x128. The backbone of the network was similar to the U-Net [27] architecture. The output of the network was the location of the nodules (x, y, z coordinates), the diameter, and a probability of being nodule. The loss function was the sum of mean squared errors, for each of the nodule location and diameter parameters, and binary cross entropy for the probability output. The training was configured with a batch size of 8, Adam as optimization algorithm, and a learning rate of 0.1 with a decay of 0.001 every 100 epochs, with a total of 450 epochs. Online hard example mining [47] with a factor of 20 times the batch size was used to optimize the network. To reduce overfitting, 3D data augmentation (such as random rotation, flip, and zoom in/out) was used during training. The network architecture is shown in Figure-2.

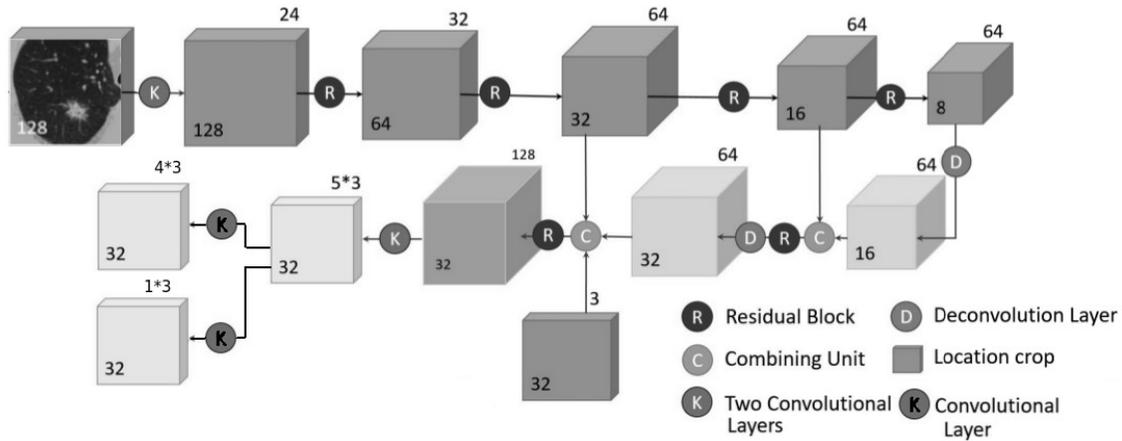

Figure-2. Architecture proposed for the nodule detection network of the pipeline.

### 3.3 Nodule re-identification

Once the nodules from two different CT scans of the same patient are detected, a second component automatically matches or re-identifies these nodules using a 3D siamese neural network (3D-SNN) as presented in [17]. A SNN [48] is made up of two components: feature extraction and classification. In the first, two subnetworks (with shared architecture and weights) process a pair of images at a time to produce two embedding feature vectors directly from the images. In the second, a head network determines whether the two embedding feature arrays are similar (i.e., correspond to the same nodule).

From the different re-identification network setups presented in [17], we used the one that obtained the best results (FIFB). This setup consisted on freezing the sibling networks of the feature extraction component with the weights of a pre-trained network, initially built for nodule identification [16]. According to [17], from the different convolution blocks of this pre-trained network, we used the output from the first block, as inputs for the head component of the 3D-SNN, since they reported better performances. The classification head component was configured with a L1-pairwise distance, a flattening layer, and a fully connected (FC) block, comprising a FC layer (with 64 units), a batch norm, a ReLU, a dropout layer and a final FC layer (with one unit). The 3D-SNNs was trained using binary cross-entropy loss function, 150 epochs, a learning rate of 1e-4, a batch size of 8, a dropout of 0.3, 10 epochs for early stopping, and Adam as the optimization algorithm. Moreover, random rotation, flip, and zoom were applied for data augmentation. Figure-3 shows the SNN architecture for the nodule re-identification problem.

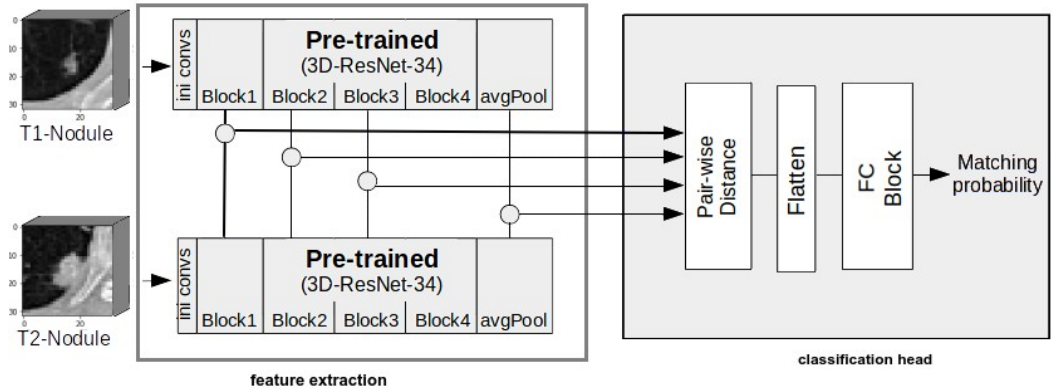

Figure 3. Architecture of the 3D-SNN for the nodule re-identification component of the pipeline.

**3.4 Nodule growth quantification**

A couple of methods are proposed for the nodule growth quantification component. The first one, already used in [17], consists on computing the diameter difference of the paired nodules from the re-identification component. The diameters of both nodules are taken from the nodule detection component of the pipeline.

The second approach differs from the previous one in that we use a probabilistic generative network (HPU) [40] to provide not only nodule growth, but also the uncertainty associated with such prediction. This type of networks is composed by two sub-networks, the prior, which models the prior distribution of possible segmentation maps for a given input image $T_i$, and the posterior, which models the joint probability distribution of the input image $T_i$ and its annotated segmentation $S_i$.

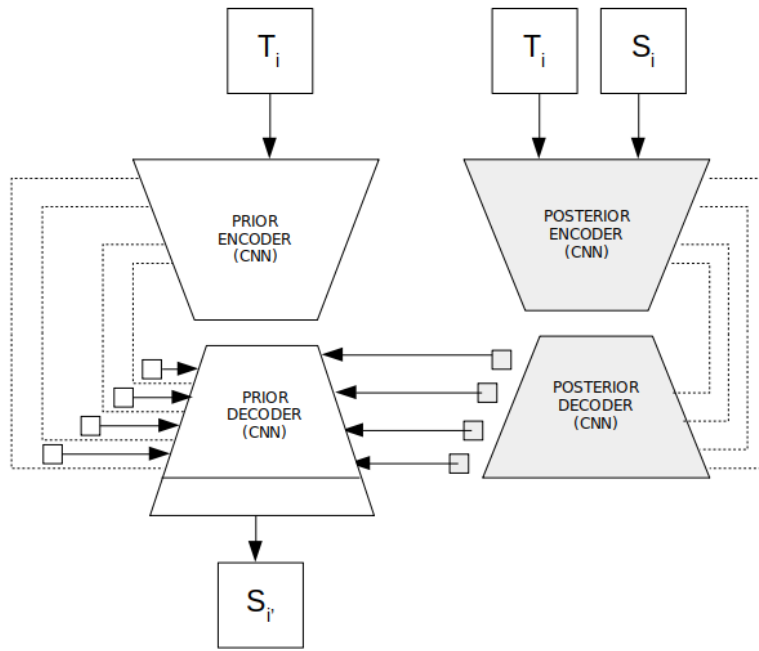

Figure-4. Hierarchical probabilistic Unet network architecture overview. $T_i$ is the nodule image i, $S_i$ is the ground truth segmentation of the nodule $T_i$, and $S_{i'}$ the predicted segmentation of the nodule $T_i$

During the training, both networks learn, in parallel, to adapt their latent distributions to be able to generate consistent segmentations. To do this, the different latent features, defined at different levels of the decoder of the posterior network, are injected to the corresponding following layer from the decoder of the prior network. In this way, gradients can flow through both networks using any stochastic gradient descent-based optimization algorithm.

At inference time, a random sample $z_i$ from the distribution defined by the different latent features, located at different levels of the decoder of the prior network, are injected into the following layer of this same network to output a new segmentation $S_{i'}$.

Following the same notations as in the original paper [40], the loss function used for training this network, also named as ELBO, is composed by the sum of the cross-entropy loss (below formulated as $P_c$) between the segmentation ground truth Y and the predicted segmentation S, given an input X and a sample z, and the distance $D_{KL}$ (Kullback-Leibler divergence) between the prior and the posterior distributions:

$$\mathcal{L}_{\text{ELBO}} = \mathbb{E}_{\mathbf{z}\sim Q}\big[-\log P_c(Y|S(X,\mathbf{z}))\big] + \beta \cdot \sum_{i=0}^{L} \mathbb{E}_{\mathbf{z}_{<i}\sim Q} D_{\text{KL}}\big(q_i(\mathbf{z}_i|\mathbf{z}_{<i},X,Y)||p_i(\mathbf{z}_i|\mathbf{z}_{<i},X)\big). \quad (1)$$

A weight factor *β* was applied to equilibrate the importance of both terms of the combined loss function (in our case we used 1 by default).

Since it is a non-deterministic network, the authors [40] proposed to evaluate this network with the generalized energy distance (GED$^2$), a metric to account for quality of the segmentation and variability in generating segmentations, according to the variability in the ground truths:

$$D_{\text{GED}}^2(P_{\text{gt}}, P_{\text{out}}) = 2\mathbb{E}\big[d(S,Y)\big] - \mathbb{E}\big[d(S,S')\big] - \mathbb{E}\big[d(Y,Y')\big], \quad (2)$$

where *d* is a distance measure (in our case, 1-IoU), S and S' are independent segmentations from the predicted distribution $P_{\text{out}}$, Y and Y' are independent segmentations from the ground truth distribution $P_{\text{gt}}$.

In the original paper, the HPU network was already trained for segmenting lung nodules using data from the LIDC dataset [49], which contains nodule segmentation annotations from up to four radiologists. However, when no nodule was marked by a radiologist, an empty segmentation image was used. This made the network to model as well the probability that there is no nodule. For our particular settings, this was undesired as the nodule detector already filters out non-nodule cases. Thus, we retrain the HPU network using the same specifications and architecture as in the original paper, but omitting empty segmentation cases.

Then, we use the HPU (prior) network to estimate the nodule growth and a measure of dispersion (standard deviation). To do this, we run N times (N=1000) the HPU network for the nodule at T1, obtaining N segmentations. From these segmentations we derive the maximum diameter of the nodule, obtaining a random vector of N diameters. We repeat this same process but for the nodule at T2. Then, we obtain the mean diameter growth as the mean difference of both random diameter vectors, and the standard deviation, as the squared root of the sum of the variances of the difference of both random diameter vectors.

### 3.5 Nodule malignancy classification

The problem of nodule malignancy classification was addressed with three different approaches, one using nodule malignancies annotated by radiologists (i.e. not confirmed cases of cancer) from a single time-point image, and the other two using nodule malignancies confirmed by diagnosis (either from biopsy or without significant growth increase during at least 2 years) with two patches of the same nodule taken at two different time-points.

The first approach consisted on re-using the best network to quantify nodule malignancy from [16]. This network (3D-CNN-MAL) receives as input a single volumetric nodule of 32x32x32. The network has a tailored architecture composed of 4 blocks of 3D CNNs, interleaved with dropout and a final dense layer with a softmax layer at the end. The outputs of this network are 3 probabilities corresponding to 3 categories of nodule malignancy (benign, suspicious and malignant).

The second approach was extracted from [18]. As in the previous approach, we selected the model with the best performance (TS-3DCNN). This model expect two volumetric input patch images of 32x32x32 centered around the nodule. The two patches correspond to the two CT scans made on the same patient but at different time-points. The network was a two-stream 3D CNN, in which two feature extraction sub-networks, with same architecture and weights, analyze in parallel the nodule patches, while the classification network part provides a cancer probability risk. Given the limited amount of longitudinal data, the siblings of the TS-3DCNN were transferred from a pre-trained 3D ResNet-34 network, used for identifying pulmonary nodules [16]. We used the features from the last layer of the second block of the 3D ResNet-34, as the ones that reported better performances. The classification head component of the TS-3DCNN was configured with a flattening, a concatenation, and a FC block layer comprising a FC layer (with 64 units), a batch norm, a ReLU, a dropout and a final FC layer (with one unit). Figure-5 shows the architecture of the TS-3DCNN network.

The third approach builds upon/integrates the other two previous approaches. Given a new nodule to classify, we predict the malignancy using 3D-CNN-MAL network and integrate this information in the TS-3DCNN network. Precisely, 6 extra features (corresponding to the 3 outcomes of the 3D-CNN-MAL for each time-point nodule) were concatenated with the features of the last fully connected layer of the TS-3DCNN.

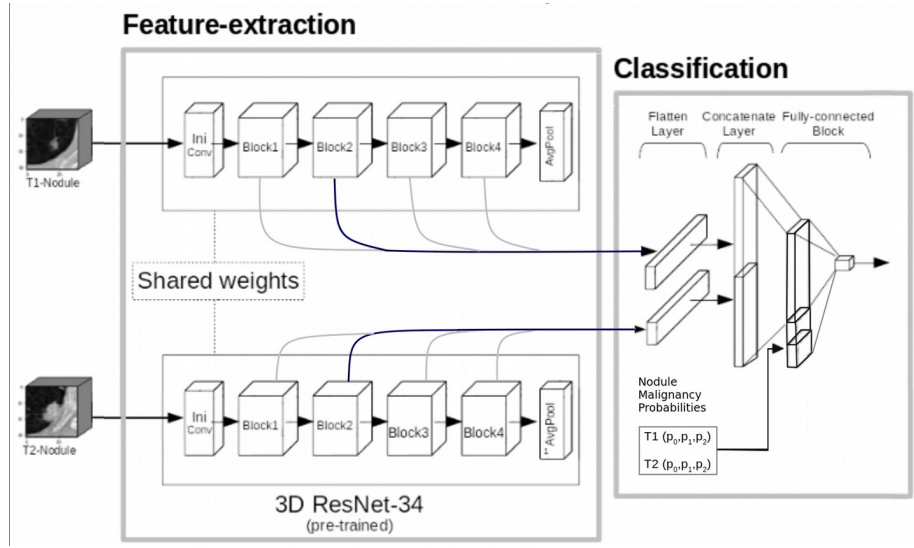

Figure-5. Nodule malignancy classification architecture of the TS-3DCNN-MAL network.

## 4. Data cohort

The complete cohort on which this study was applied contains 151 patients for whom at least two chest CT scans were taken at two different time points (T1 and T2). This cohort (VHLung) was collected from the Vall d'Hebron University Hospital (Barcelona). For each patient, two different radiologists detected and matched the most relevant nodule on the two CT scans, providing its locations and diameters on each scan.

The ethics approval for this study was obtained from the Medication Research Ethics Committee of Vall d'Hebron University Hospital (Barcelona) with reference number PR(AG)111/2019 presented on 01/03/2019. Inclusion criteria were patients without a previous neoplasia, with a confirmed diagnosis, and with visible nodules (>= 5 mm) in at least two consecutive CT scans.

### 4.1 Scaners and protocols

The chest helical CT studies were performed using different scanners: Phillips (Brilliance 16/64, iCT 256), Siemens (SOMATOM Perspective/ Definition) and General Electrics (LightSpeed16). Acquisition and reconstruction protocols were set according to subject biometrics and clinical inquiry: 100–120 kV, 33-196 mAs and exposure time 439-1170 ms. Each image had 512x512 pixels with 16-bit gray resolution, spacing between slices 0.75-1.5 mm and slice thickness 1-5 mm.

**4.2 Data**

To evaluate the pipeline, we randomly selected 38 patients from the complete cohort (the remaining data were used for training purposes). This test set was composed by 25 cancer and 13 benign cases. Also, the interquantile interval of time between current and previous CT examinations of this test set ranged from 288 to 775 days. The mean size of the annotated test set nodules was 10.73 +/- 3.80 mm at T1 and 13.62 +/- 4.94 mm at T2, and their mean growth size was 2.89 +/- 4.35 mm. Figure-6 provides a visual description of the test dataset.

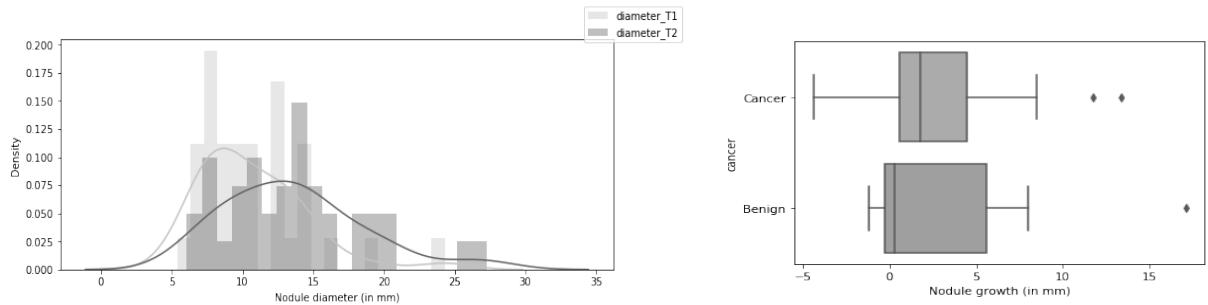

Figure 6. Histogram and boxplot of the nodule size of the VHLung test set .

## 5. Results

In this section, we report the performance of the pipeline obtained on the VHLung test set, broken down into each of its components.

**5.1 Nodule detection**

First of all, we evaluate the ability of the pipeline to detect the annotated nodules (one per each CT) among all nodules predicted by the 3D-FasterCNN network. Therefore, we measured the performance of the network to find the annotated nodules in the least number of predicted nodule candidates. Results show that, taking a reasonable threshold of top-32 predicted nodule candidates per CT, the pipeline obtained a sensitivity score of 0.973 on the 76 CT scans of the test set (taken individually), missing only 2 of them. Figure-7 shows the sensitivity performances using different amounts of predicted nodule candidates, obtained from 10 random subsets with replacement for both the training and test sets.

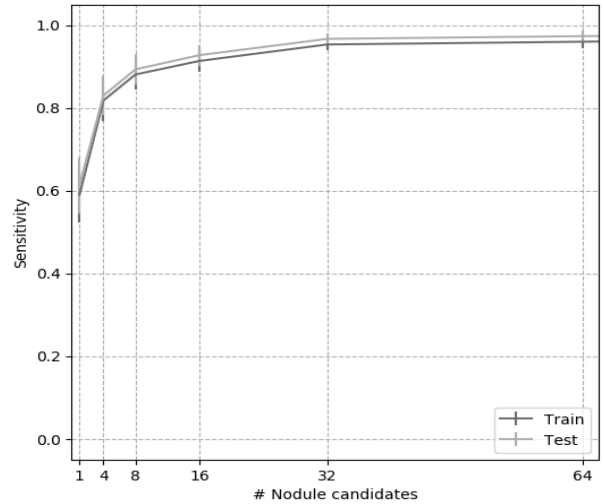

Figure-7. FROC-curve of the malignant nodule detection algorithm for training and test partition.

## 5.2 Nodule re-identification

To evaluate the second stage of the pipeline, we used the 36 pairs of CT scans where the pipeline found the radiologist's annotated nodules. For each pair of CTs, we input 64 (32 per each time-point) nodule candidates into the 3D-SNN network to obtain those that correspond. In total, the 3D-SNN network reported only 4 CT-pairs incorrectly matched with an accuracy of 0.888. Table-1 provides a summary of the results for the re-identification step, stratifying them by the initial size of the nodules.

|  | Small | Medium | Large | Total |
|---|---|---|---|---|
| Accuracy | 1.0 | 0.84 | 0.75 | 0.888 |
| #Nodule-pairs | 13 | 19 | 4 | 36 |

Table-1. Performance of the re-identification component of the pipeline.

## 5.3 Nodule growth quantification

Two different approaches for nodule growth quantification were evaluated on the 32 matching nodules obtained from the nodule re-identification component. The first approach used the predicted nodule diameter measurements from the 3D-FasterCNN network, while the second used the predicted nodule diameter measurements from the HPU network. For a proper usage of the HPU in the context of the nodule growth quantification, we retrain this network according to [40] with same data from LIDC dataset but omitting "empty" cases where radiologists did not mark any nodule in the axial slices. The model reported a $GED^2$ of 0.38 and a reconstruction Dice of 0.91.

Table-2 shows the mean absolute error, mean squared error and r-coefficient of correlation with respect to the ground truth. Results show the mean and 2 standard error associated with 95% of confidence, obtained with a 1000 bootstraps with replacement of the test set. Figure-8 shows estimated nodule growth sizes from both networks per each nodule of the test set.

|  | MAE | MSE | $R^2$ |
| --- | --- | --- | --- |
| 3D-FasterCNN | 1.400 +/- 0.422 | 3.333 +/- 1.927 | 0.637 +/- 0.344 |
| **HPU** | **1.348 +/- 0.370** | **2.889 +/- 1.561** | **0.667 +/- 0.418** |

Table-2. Nodule growth performance comparison between FasterCNN and HPU networks.

These results show that the HPU network (segmentation based approach) provides closer estimates to radiologists annotations than the 3D-FasterCNN network. For the HPU network we also evaluated how relatively far the estimated nodule growth distribution was respect to the radiologist ground truth. To do this, we computed the Mahalanobis distance (and its associated probability) between the radiologist nodule growth annotation and the center of the estimated diameter growth distribution. Additionally, we computed the probability of being the radiologist annotations, 1 or 2 standard deviation away from the center of the estimated nodule growth distribution. Table-3 summarizes these results, in which the values represent the mean and 2 standard errors associated with a 95% of confidence, obtained from 1000 bootstraps with replacement of the test set.

|  | Test |
|---|---|
| Mahalanobis distance (RX_growth, distribution) | 0.383 +/- 0.205 |
| P(RX_growth) <= 1 std to the mean (of the distribution) | 0.874 +/- 0.117 |
| P(RX_growth) <= 2 std to the mean (of the distribution) | 1.0 +/- 0.0 |

Table-3. HPU nodule growth uncertainty performance

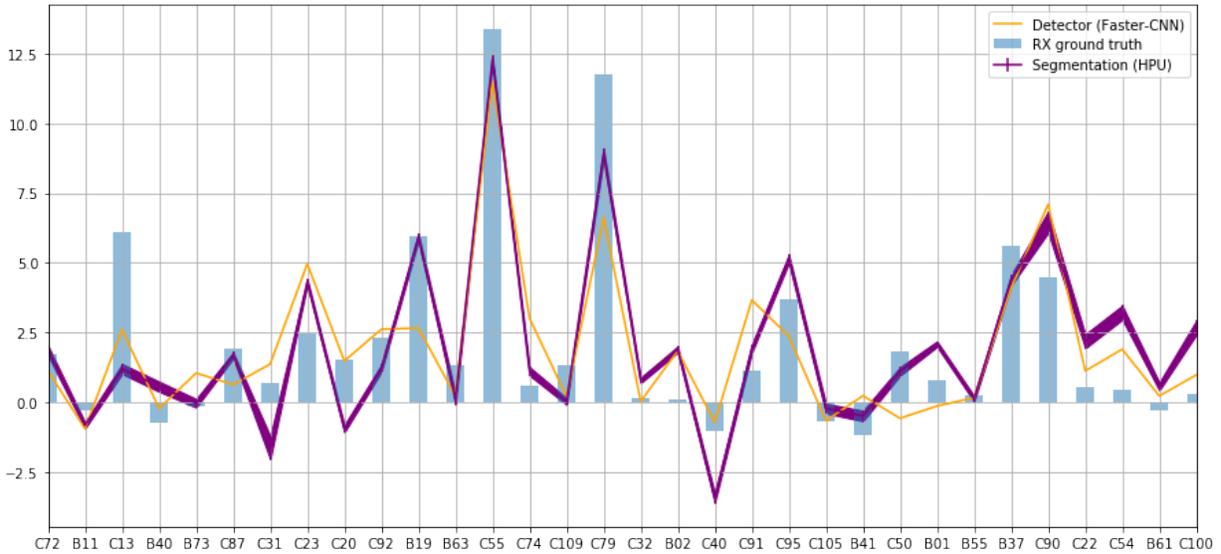

Figure-8. Comparative of radiologist growth measurements with results from nodule detector and nodule segmentation

## 5.4 Nodule malignancy classification

Three different methods (3DCNN-MAL, TS-3DCNN and TS-3DCNN-MAL) were evaluated on the resulting 32 matching lung nodules (65% of them cancerous) from the re-identification step. For the evaluation of these methods, we directly used the 3DCNN-MAL classifier on the evaluation cases, while for the other two classifiers, to avoid data leakage for this evaluation, we retrained them before being evaluated, using the training partition of the VHLung with a 10-fold cross-validation. Particularly for the 3DCNN-MAL classifier, as it outputs 3 probabilities, we assumed cancer prediction when this classifier reported as maximum probability either the category suspicious or malignancy.

Table-4 shows the resulting classification performances for these models on the 32 matching nodules. We computed, precision (PREC), recall (REC) and specificity (SPEC), as well as balanced accuracy (BA). Due to the data was unbalanced towards the cancer case, the BA was used as the reference metric. Reported values in Table-4 are the mean and 2 standard errors (associated with a 95% of confidence).

|  | TEST | | | |
|---|---|---|---|---|
|  | **BA** | **PREC** | **REC** | **SPEC** |
| 3DCNN-MAL | 0.776+/-0.153 | 0.808+/-0.152 | 1.0+/-0.0 | 0.552+/-0.307 |
| TS-3DCNN | 0.810+/-0.203 | 0.899+/-0.167 | 0.791+/-0.309 | 0.829+/-0.296 |
| **TS-3DCNN-MAL** | 0.825+/-0.201 | 0.910+/-0.179 | 0.821+/-0.33 | 0.830+/-0.348 |

Table-4. Performance of the different nodule malignancy classifiers of component of the pipeline.

The 3DCNN-MAL classifier obtained a balanced accuracy score of 0.77, while the TS-3DCNN achieved a 0.81. However, the TS-3DCNN-MAL, which integrated the outcomes of the 3DCNN-MAL model, improved the balanced accuracy score of the TS-3DCNN model, a 1.5%. For further comparison of these models, we show Figure-9 with the ROC-curves of the two best models.

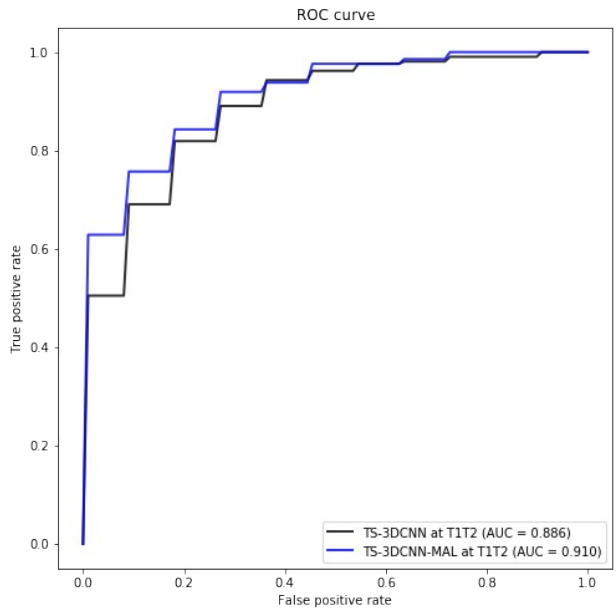

Figure 9. Roc curves of the nodule malignancy classification models

For a further intuition and visual interpretability of the areas of the images that the TS-3DCNN-MAL network took most seriously in deciding which class to assign to the image, we extracted the Grad-CAMs features for this classifier [50]. In particular, on the last layer of the first block of the TS-3DCNN-MAL network we obtained the gradients and the feature activations, and they were multiplied after being pooled on the channel dimension. Figures-10,11 shows the results of this visualization technique on the three-planes of different lung nodules (either benign as malign).

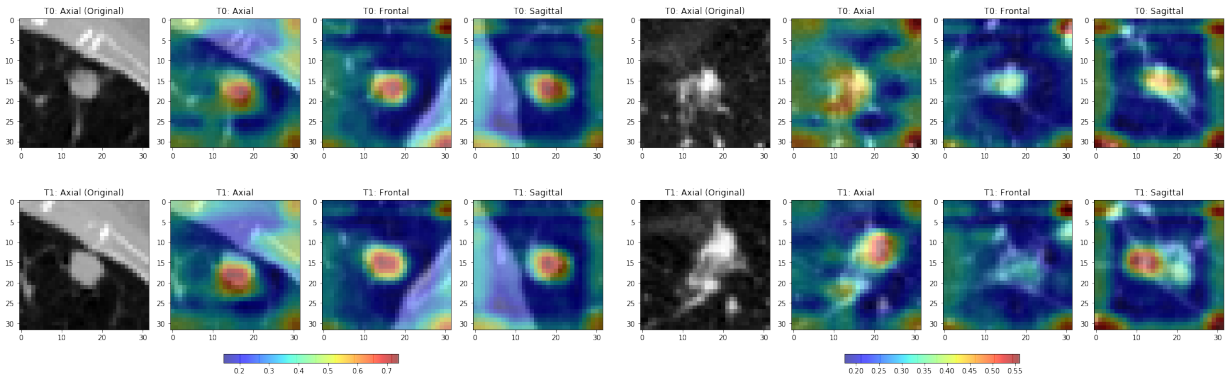

Figure-10. Grad-CAM features from the TS-3DCNN-TL-MAL network for 2 malign nodules

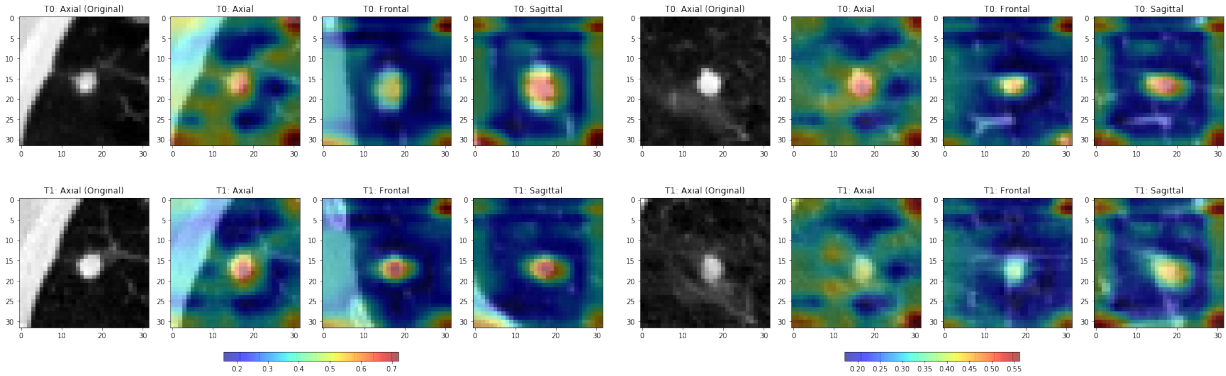

Figure-11. Grad-CAM features from the TS-3DCNN-TL-MAL network for 2 benign nodules

## 6. Discussion

The still incomplete knowledge of malignant patterns in the course of multifactorial diseases such as lung cancer makes it necessary to support physicians with automatic, fast and reliable predictive tools to reduce the workload in radiological services. Unfortunately, the vast majority of research is still focused on specific tasks with data from single time-points [16,22-24,41-43], which limits their potential impact and usability in real clinical settings.

In this chapter, we presented a computer vision pipeline aimed at automatizing the main tasks involved in the lung cancer follow-up. To do this, we relied on a deep learning solution aiming at modeling the temporal evolution of this disease to assess nodule growth and malignancy.

The pipeline was formed by 4 different components: nodules detection, nodule re-identification, nodule growth and nodule malignancy classification. The evaluation of the pipeline was done on an independent test set composed by 38 CT pairs. To train the models, we used two different training datasets, LIDC [49] for nodule detection and nodule growth estimation, and VHLung for nodule re-identification and cancer classification.

The evaluation of the first component of the pipeline, aimed at detecting the suspicious nodules marked by the radiologists. To this end, we re-used a deep convolutional network based on the 3D Faster-CNN scheme already published in a previous work [17]. This network reported on the available data for the evaluation of the pipeline (test set of the VHLung dataset) a sensitivity score of 97.6% for 32 nodule candidates. When evaluating this model in a larger test set (e.g. LIDC test set), the performance decreased to 84% sensitivity at 1 false positive. This value is slightly below compared with top performances (81.7% at 0.125 FP) in LUNA-16[1] benchmarks. However, it is not clear if those performances are realistic or just an overfit on the provided dataset.

For the nodule re-identification component of the pipeline, we re-used the best model reported in [17]. One of the main benefits of using this approach for matching nodules is that no previous registration of the lung CTs was required (which usually it is slow and introduce artifacts in the original images). The results for nodule re-identification reported also high performances (88.8% of accuracy). However, we should note that they are still a bit below the performances reported, by the same method, when performing in an isolated way, the matching task (92% of accuracy).

Two different approaches were proposed and compared for the nodule growth component. The first one relied directly on the predicted nodule diameters reported by the model of the nodule detection component of the pipeline. Therefore the growth was computed from the substraction of both measurements. The second approach for nodule growth estimation relied on a hierarchical probabilistic U-Net (HPU) [40]. This method, based on the generation of several feasible segmentations of the nodule, allowed us to provide an estimation of the growth of the nodule together with an uncertainty of the reliability of the model on this measure. This method obtained the best result, specifically, a mean absolute error (MAE) of 1.34 mm (with an standard error of +/-0.37 mm at 95% of confidence). This approach slightly outperformed by 0.05 mm of MAE the previous approach based on the nodule detection network. Somehow this result was expected since the nodule detection network was trained without any information regarding the contour of the nodules. Nonetheless, both approaches reported errors that were below 2 mm, the threshold determined by radiological guidelines from which to consider nodule growth [38]. Beyond these results, the ability of the HPU-based approach to provide a measure of uncertainty could help clinicians to make better decisions since it provides how confident the model is about its predictions.

---

1  https://luna16.grand-challenge.org/

Regarding lung cancer classification, our best model (TS-3DCNN-MAL) obtained 82.3% of balanced accuracy score. This performance is competitive with those reported from recent cancer classification systems. For instance, in [42] they achieved 86% and 87% of precision and recall, while we obtained 91% and 81.9%. In [44] , they reported an AUC of score 92.6% while our model obtained 91.1%.

Despite the notable performances of the pipeline, our work still presents several limitations. First, the great heterogeneity and complexity of the problem makes the amount of data used for the evaluation of the pipeline too small. Hence, greater emphasis is needed on collecting new data for a more comprehensive evaluation. Second, although in the clinical practice the nodule growth is measured with the size of the diameter, we believe building a growth detection method relying on 3D volumetric measures should capture more accurately the patterns of nodule growth. Third, more efforts on visualization and explanaibility techniques could be done to abilitate more transparency in the models of the pipeline and thus an easier implantation of this tool in clinical domains.

Finally, several future works can be envisaged. The integration of non-image data, such as the clinical history of the patient, could be an added-value on the pipeline for modeling the whole context of the disease. Also, further efforts in fine-tuning the current networks or adopting recent advances in computer vision [54,55] could lead to an overall improvement of the performances reported.

## 7. Conclusion

In this chapter we address the problem of supporting radiologists in the longitudinal management of lung cancer. Therefore, we proposed a deep learning pipeline, composed of four stages that completely automatized from the detection of nodules to the classification of cancer, through the detection of growth in the nodules. In addition, the pipeline integrated a novel approach for nodule growth detection, which relied on a recent hierarchical probabilistc U-Net adapted to report uncertainty estimates. Also, a second novel method was introduced for lung cancer nodule classification, integrating into a two stream 3D-CNN network the estimated nodule malignancy probabilities derived from a pre-trained nodule malignancy network. The pipeline was evaluated in a longitudinal cohort and reported comparable performances to the state of art.